\documentclass[aps,prb,twocolumn,groupedaddress,amsmath,amssymb,amsfonts,graphics,mathrsfs]{revtex4}
\usepackage[mathscr]{euscript}
\usepackage{feynmf}
\usepackage{bm}
\usepackage{wasysym}
\usepackage{ulem}
\usepackage{graphicx}
\usepackage{bm}
\usepackage{color}
\usepackage{verbatim}

\def\<{\begin{equation}}
\def\>{\end{equation}}

\begin{document}



\title{RKKY in half-filled bipartite lattices: graphene as an example}

\author{Saeed Saremi}
\affiliation{Department of Physics, Massachusetts Institute of Technology,
Cambridge, Massachusetts 02139}

\date{\today}

\begin{abstract}
We first present a simple proof that for any bipartite lattice at half filling the RKKY interaction is antiferromagnetic between impurities on opposite (i.e., A and B) sublattices and is ferromagnetic between impurities on the same sublattices. This result is valid on all length scales. We then focus on the honeycomb lattice and examine the theorem in the long distance limit by performing the low energy calculation using Dirac electrons. To find the universal (cutoff free) result we perform the calculation in  smooth cutoff schemes, as we show that the calculation based on a sharp cutoff leads to wrong results. We also find the long distance behavior of the RKKY interaction between ``plaquette" impurities in both coherent and incoherent regimes. \end{abstract}

\maketitle

\begin{fmffile}{fmfrkky}

\section{Introduction}
The RKKY interaction between magnetic impurities in a metallic environment plays a crucial role in the way magnetic impurities order both in the dilute and the Kondo lattice limits.\cite{Doniach} Perfect nesting of the Fermi surface provides a mechanism for ordering, due to the divergence of the spin susceptibility at the nesting wavevector.\cite{Auerbach} The square lattice at half filling provides the best known example for perfect nesting and the consequent $(\pi,\pi)$ antiferromagnetic ordering of the magnetic impurities. The {\it minimal} form of nesting appears in the undoped single layer of graphite\cite{Novoselovetal} where the Fermi surface shrinks to two Dirac points. However we show that the particle-hole symmetry provides another mechanism for ordering of the magnetic impurities in a metallic environment on {\it all} length scales. This result can be applied to the undoped graphene and is of value since low energy calculations are only valid in the long distance limit. Furthermore, this result provides a test for the low energy calculations and  we show that the calculation based on a {\it sharp} cutoff fails this test. 
 We also study the RKKY interaction between localized spins with more complicated Kondo interactions. Interestingly, the behavior of the RKKY interaction is qualitatively different for different types of magnetic impurities. It will be interesting to see which one  of the scenarios  we considered -- if not all -- can be realized in the lab.

The organization of the paper is as follows. In Sec. \ref{SEC:lattice} we consider the simplest Kondo perturbation, where the impurity is localized at a lattice site, and only has an \textit{on-site} Kondo interaction with the conduction electron spin. We prove a theorem for the RKKY interaction between these ``site impurities" in bipartite lattices  with hopping between opposite AB sublattices at half filling. The result is that the sign of the RKKY interaction  depends \textit{only} on whether the impurities are localized at opposite sublattices (antiferromagnetic) or on the same sublattices (ferromagnetic). The sign is dictated by particle-hole symmetry and is thus valid on all length scales.

Section \ref{sec:low-energy} focuses on the half-filled honeycomb lattice with nearest neighbor hopping, and examines the above theorem in the long distance limit. We do a low energy calculation for the RKKY exchange in different cutoff schemes to obtain the long distance behavior of the exchange. As will be explained, due to the nature of the singularity of the spin susceptibility near the Dirac nodes,  the use of sharp cutoff is inappropriate. We do the sharp cutoff calculation to demonstrate this fact. We then examine two smooth cutoffs to find the long distance behavior of the exchange. \textit{Two} cutoffs are considered to make sure that the calculation is cutoff independent. They both lead to the same answer in the long distance limit. 

It is important to note that although both antiferromagnetic and ferromagnetic exchange have the same algebraic decay power in the long distance limit, the coefficient of their decays are not the same. Furthermore, focusing on the RKKY in a single class (i.e. the same or opposite sublattices), there is also magnitude ``oscillations" of the form $\cos\left(2 {\mathbf k}_D\cdot ( {\bm R}-{\bm R'} )\right)$ due to the non-analyticity of the susceptibility at the wavevector connecting the two nodes.

In Sec. \ref{SEC:plaq} we discuss the extension of the RKKY between site impurities 
to more general cases. In particular, we find the qualitative behavior of the RKKY for magnetic impurities sitting at the center of the hexagons of the honeycomb lattice. 
We distinguish this type of impurity by calling it \textit{plaquette} impurity (vs.  site impurity). In writing the Kondo perturbation, these plaquette impurities can couple coherently or incoherently with the conduction electrons. We find that, for incoherent Kondo couplings, the RKKY exchange between  plaquette impurities is \textit{ always antiferromagnetic}. 

We then obtain the RKKY behavior for plaquette impurities when they have a \textit{coherent} Kondo interaction with the conduction electrons around the plaquette. We find that, due to a nontrivial phase cancelation, the $1/R^3$ algebraic tail, present in all other situations we considered, vanishes.

We conclude by summarizing our results and mentioning our disagreements -- specially in the \textit{sign}  of the RKKY interaction -- with other attempts\cite{VLSG05,DLB06} made in calculating the RKKY interaction in graphene. We also discuss the direct implication of this paper for the Kondo lattice model on the honeycomb lattice.

\section{lattice results \label{SEC:lattice}}
To establish notation we quickly review the RKKY interaction.\cite{Kittel} Imagine putting 2 test localized spins $S$ at lattice sites i and j $\rm  (i \neq j)$ and assume they have a small on-site spin exchange interaction  with the conduction electrons spin $s$. This perturbs the free hopping Hamiltonian 
\begin{equation} \label{EQ:Hamiltonian}
\hat{H} = -t \sum_{\rm \left<ij\right>,a} \left( \hat{c}_{\rm ia}^{\dagger} \hat{c}_{\rm ja} + {\rm H.c.} \right)
\end{equation}
by 
\begin{equation} \label{EQ:pert}
\delta \hat{H} = \mathfrak{j}(\bm{S}_{\rm i}\cdot \bm{s}_{\rm i} + \bm{S}_{\rm j}\cdot \bm{s}_{\rm j} ) \ \ \ (|\mathfrak{j}| \ll t).
\end{equation}
In perturbation theory 
 the leading interaction induced by this term is
\begin{equation} 
\hat{H}_{\rm RKKY} = J_{\rm ij} \bm{S}_{\rm i}\cdot \bm{S}_{\rm j},
\end{equation}
where $J_{\rm ij}$ is given by
\begin{equation} \label{EQ:Jij}
J_{\rm ij} = -\mathfrak{j}^2\int d\tau~ \bigl<s^-_{\rm i}(\tau) s^+_{\rm j}(0)\bigr>.
\end{equation}
To be precise $J_{\rm ij}$ is the coefficient of the $S_{\rm i}^+S_{\rm j}^-$. However due to the SU(2) flavor symmetry of the unperturbed Hamiltonian; $J_{\rm ij}^{-+}$, $J_{\rm ij}^{+-}$ and $J_{\rm ij}^{zz}$ are all equal. It is also understood that this is the static (imaginary-time-averaged) part of the RKKY interaction.

Nearest neighbor hopping on a bipartite lattice can be viewed as a special case of hopping between A and B sublattices. The particle-hole transformation
\begin{align}\label{EQ:ph}
c_{\rm i a}(\tau) \rightarrow (-1)^{\rm i} ~\bar{c}_{\rm i a}(\tau),
\end{align}
where $c$ and its conjugate $\bar{c}$ are the Grassmann fields, leaves the Lagrangian invariant. Lattice sites are labeled such that i is odd on A sublattices and even on B sublattices. This  transformation sends the chemical potential $\mu$ to $-\mu$. However at half filling $\mu=0$ and the particle-hole transformation is a symmetry of  the partition function. 

Using Wick's theorem we have
\begin{equation} \label{EQ:GijGji}
\begin{split}
J_{\rm ij} &= - \mathfrak{j}^2 \int d\tau~ \bigl< \bar{c}_{\rm i \downarrow}(\tau) c_{\rm j \downarrow}(0)\bigr> \bigl< c_{\rm i \uparrow}(\tau) \bar{c}_{\rm j \uparrow}(0)  \bigr>.
\end{split}
\end{equation}
We drop the spin indices from now on since the Green's functions do not depend on the spin flavor. Since the ground state is particle-hole symmetric, the particle-hole symmetry of the Lagrangian  implies  
\begin{equation}
\bigl< c_{\rm i}(\tau) \bar{c}_{\rm j}(0)  \bigr> = (-1)^{{\rm i+j}} \bigl< \bar{c}_{\rm i}(\tau) c_{\rm j}(0)\bigr>.
\end{equation}

The above relation immediately results in 
\begin{equation} 
J_{\rm ij} = -\frak{j}^2~(-1)^{\rm i+j}  \int d\tau~  G_{\rm ij}(\tau)^2,
\end{equation}
where $G_{\rm ij}(\tau)$ is defined to be
\begin{equation}
G_{\rm ij}(\tau) = \bigl< \bar{c}_{\rm i}(\tau) c_{\rm j}(0)\bigr>.
\end{equation}
In addition $G_{\rm ij}(\tau)$ is real, as the space-time matrix connecting  the Grassmann variables is real.
Therefore
\begin{equation}\label{EQ:AF}
\frac{J_{\rm ij}}{|J_{\rm ij}|} = (-1)^{\rm i+j+1}.
\end{equation}

The same result is obtained if the Hamiltonian of Eq.~(\ref{EQ:Hamiltonian}) contains more general hopping terms, but only between AB sublattices. In summary, for any bipartite lattice at half filling, with hopping only between AB sublattices, the RKKY interaction  is antiferromagnetic between impurities on opposite sublattices and is ferromagnetic between impurities on the same sublattices.


\section{low energy calculation \label{sec:low-energy}}
Next we focus on the honeycomb lattice with nearest neighbor hopping and examine the theorem we proved in the long distance limit. 
The low energy theory is known to be governed by a 2+1 dimensional Dirac action containing the Dirac spinors -- with internal A-B flavor -- residing near the two independent nodes $\pm \bm{k}_D$. 
Being ``near" the nodes is formulated by introducing a momentum cutoff and a cutoff scheme in the calculations.

The division of the honeycomb lattice to A and B sublattices is a necessity, as the Bravais lattice has a basis with two sites. Let us set the distance  between the nearest neighbor sites to be 1. In the limit $|{\bm R}-{\bm R'}|\gg 1$, the RKKY interaction $J_{\rm AB}({\bm R}-{\bm R'})$ is calculated by Fourier transforming Eq.~(\ref{EQ:GijGji}) and constraining the momentums to be near the Dirac nodes:
\begin{widetext}
\begin{equation}\label{EQ:JAB}
J_{\rm AB}({\bm R}-{\bm R'}) \approx -\mathfrak{j}^2 \sum_{\rm{D,D'}} \int \frac{d^2{\bm q}}{(2\pi)^2}e^{i\bm{q}\cdot({\bm R}-{\bm R'})} e^{i\rm{(D-D')}\mathbf{k}_D\cdot({\mathbf {R-R'}})}   \chi^{\rm{DD'}}_{\rm AB}(q_0=0,{\bm q}) \mathcal{C}_\Lambda(|\bm{q}|).
\end{equation}
\end{widetext}

Here $\bm{R}$ and $\bm{R'}$ refer to the Bravais lattice vectors.  D and D$'$ are either $+1$ or $-1$. They denote near which Dirac nodes $\pm \bm{k}_D$ the (space) momentums reside. 
$\Lambda \ll 1$ is the cutoff  and $\mathcal{C}_\Lambda(|\bm{q}|)$ is a function that takes care of cutting off the momentum.  Three examples for $\mathcal{C}_\Lambda(|\bm{q}|)$ will be given below. 

$\chi^{\rm{DD'}}_{\rm AB}(q)$ is given diagrammatically by
\\
\\
\begin{equation} \label{EQ:chi}
\chi^{\rm{DD'}}_{\rm AB}(q)=\int\frac{d^3 k}{(2\pi)^3}\ \ 
{\rm A}\ \parbox{11mm}{
\begin{fmfgraph*}(10,6)
\fmfleft{i}
\fmfright{o}
\fmf{dbl_plain_arrow,label=$k+q; \rm{D}$,right}{o,i}
\fmf{dbl_plain_arrow,label=$k; \rm{D'} $,right}{i,o}
\fmfv{
decor.shape=cross,decor.filled=full,decor.size=2thick}{i}
\fmfv{
decor.shape=cross,decor.filled=full,decor.size=2thick}{o}
\end{fmfgraph*}
}{\rm B}\ .
\end{equation}
\\
\\

We use the continuum limit ``translation" of the fields in terms of microscopic variables provided by Eq. (45) and (50) of Ref.~\onlinecite{saeed07} to obtain:
\begin{align}
\chi_{\rm AB}^{++}(0,{\bm q})&=\frac{3|{\bm q}|}{64},\\
\chi_{\rm AB}^{--}(0,{\bm q})&=\chi_{AB}^{++}({\bm q}),\\
\chi_{\rm AB}^{+-}(0,{\bm q}) &= \frac{1}{64|{\bm q}|}\left(q_x+iq_y\right)^2, \\
\chi_{\rm AB}^{-+}(0,{\bm q}) &= \chi_{AB}^{+-}({\bm q})^*.
\end{align}
In the above calculations the Dirac dispersion velocity $v_c$ is set to be 1 by scaling the space dimensions.

To calculate $J_{\rm AB}({\bm R}-{\bm R'}) $ we examine the following cutoff functions :
\begin{align}
\mathcal{C}_\Lambda^1 (|\bm{q}|) &= \theta(\Lambda-|\bm{q}|),\\
\label{EQ:C2} \mathcal{C}_\Lambda^2 (|\bm{q}|) &= e^{-|\bm{q}|/\Lambda},\\ 
\label{EQ:C3} \mathcal{C}_\Lambda^3 (|\bm{q}|) &= e^{-\bm{q}^2/\Lambda^2}, 
\end{align}
where $\theta$ is the step function.

The issue is that the susceptibility is dominated by larger values of $|\bm{q}|$, and the decay that phase integration causes, in the sharp cutoff scheme, is not strong enough to compensate that. This issue becomes more severe in the continuum limit $\Lambda |{\bm R}-{\bm R'}|  \rightarrow \infty$. 

Since small $|\bm{q}|$ \textit{must} dominate the large distance behavior, we should adopt a different cutoff scheme than the sharp cutoff.  For the same reason, the answer should be universal, not just the power of decay, but also its coefficient. That is why we examine two different cutoff schemes $\mathcal{C}_\Lambda^2$ and $\mathcal{C}_\Lambda^3$ to make sure that we find the cutoff independent answer.

\subsection{Examining different cutoffs}

Next we do the calculations for the three cutoff schemes we have considered. We first show explicitly that the sharp cutoff is inappropriate. We then find the universal answer for the RKKY interaction using the smooth cutoffs $\mathcal{C}_\Lambda^2$ and $\mathcal{C}_\Lambda^3$.

\subsubsection{sharp cutoff}

Doing the integral of Eq.~(\ref{EQ:JAB})  for $\rm D=D'=+1$ using $\mathcal{C}_\Lambda^1$ results in 
\begin{equation}
J_{\rm AB}^{++}({\bm R}-{\bm R'}) \propto \frac{1}{|{\bm R}-{\bm R'}|^3}\int_0^{\Lambda |{\bm R}-{\bm R'}|} dx~x^2 J_0(x).
\end{equation}
The Bessel function $J_0(x)$ is obtained by doing the angular integration. The integrand in the above equation is a widely oscillating function in the limit $\Lambda |{\bm R}-{\bm R'}| \gg 1$. Using the asymptotic form for $J_0(x)$  one easily obtains
\begin{equation} J_{\rm AB}^{++}({\bm R}-{\bm R'}) \propto \frac{1}{|{\bm R}-{\bm R'}|^{3/2}} \sin(\Lambda |{\bm R}-{\bm R'}| - \pi/4).\end{equation}
 The {\it changed scaling form}, as well as the sine oscillations, are generated by the sharp cutoff in momentum space.  The sine causes sign oscillations in contradiction to the theorem we proved in the first section, thus making the sharp cutoff inappropriate.
 Again even without the theorem; we can see that the integral is dominated by larger values of $|\bm{q}|$, and that is enough to make the sharp cutoff inappropriate.

\subsubsection{smooth cutoffs}

Because of this issue we use the smooth cutoffs given by Eq.~(\ref{EQ:C2}) and (\ref{EQ:C3}). As expected, $\mathcal{C}_\Lambda^2$ and  $\mathcal{C}_\Lambda^3$,  both lead to the same  result in the continuum limit.  The few integrals needed for this calculation are given in the Appendix \ref{APP:integrals}. 
The final result is
\begin{equation} \label{EQ:J_AB}
J_{\rm AB}({\bm R}-{\bm R'}) \approx \frac{3\mathfrak{j}^2}{64\pi }\frac{1+\cos\bigl(2 {\mathbf k}_D\cdot ( {\bm R}-{\bm R'} )\bigr)}{ |{\bm R}-{\bm R'}| ^3},
\end{equation}
where $\approx$ is used since it is the continuum limit $\Lambda |{\bm R}-{\bm R'}| \rightarrow \infty$ result.
We also mention the results for $\chi_{\rm AA}^{\rm D D'}$:
\begin{equation}
\chi_{\rm AA}^{\rm D D'}(0,{\bm q}) =-\frac{|{\bm q}|}{64}.
\end{equation}
The same cutoff functions $\mathcal{C}_\Lambda^2$ and $\mathcal{C}_\Lambda^3$ results in 
\begin{equation}  \label{EQ:J_AA}
J_{\rm AA}({\bm R}-{\bm R'}) \approx \frac{-\mathfrak{j}^2}{64\pi }\frac{1+\cos\bigl(2 {\mathbf k}_D\cdot ( {\bm R}-{\bm R'} )\bigr)}{ |{\bm R}-{\bm R'}| ^3}.
\end{equation}
 
\subsection{RKKY for plaquette impurities \label{SEC:plaq}}

The perturbation we considered in Eq.~(\ref{EQ:pert}) is the simplest Kondo perturbation one can consider. In general the localized spin could have interactions with several conduction electrons. In the honeycomb lattice this can be realized experimentally by having the localized spins near the center of the graphene's hexagons. 

To study these situations, let us  consider a slight change of notations  in denoting the localized spins. The Greek alphabet index $\bm{S}^{\alpha}$ is used to denote the localized spins in these situations. This is in contrast to the notation $\bm{S}_{\rm i}$ for the \textit{on-site} perturbations we started this paper with. To establish notation we first consider the simpler theoretical problem in which the impurity has an incoherent Kondo interaction with the conduction electron spins around the plaquette. The perturbation to the free hopping Hamiltonian is then

\begin{equation} \label{EQ:general_pert}
\delta \hat{H} = \mathfrak{j}~ T^{\alpha}_{\rm i} \bm{S}^{\alpha}\cdot \bm{s}_{\rm i} \ \ \ (|\mathfrak{j}| \ll t),
\end{equation}
where the sum over $\alpha$ (localized spins) and i (lattice sites) are understood. We assume $\bm{S}^{\alpha}$ has interactions with only a few $\bm{s}_{\rm i}$'s. For example, if $\bm{S}^{\alpha}$ is located near the center of a hexagon in the honeycomb lattice, we assume $T^{\alpha}_{\rm i}$ has only 6 nonzero elements, corresponding to the sites surrounding the hexagon.

The RKKY interaction between $\bm{S}^{\alpha}$ and $\bm{S}^{\beta}$ is
\begin{equation} 
J^{\alpha \beta} \bm{S}^{\alpha}\cdot \bm{S}^{\beta},
\end{equation}
where $J^{\alpha \beta}$ is given by
\begin{equation}
J^{\alpha \beta} = T^{\alpha}_{\rm i}T^{\beta}_{\rm j} J_{\rm ij},
\end{equation}
and $J_{\rm ij}$ is given by Eq.~(\ref{EQ:Jij}). 

To be concrete let us consider the honeycomb lattice and imagine that the localized spins are located at the center of the hexagons and are widely separated. 
We consider the ``s-wave" model in which  the non-zero elements of  $T^{\alpha}_{\rm i}$ are all 1. $J^{\alpha \beta}({\bm R}-{\bm R'})$ is then given by summing over 36 terms that are given by either Eq.~(\ref{EQ:J_AB}) or Eq.~(\ref{EQ:J_AA}). Since A and B sites surrounding the hexagons belong to different Bravais lattice sites, the separations in  Eq.~(\ref{EQ:J_AB}) or Eq.~(\ref{EQ:J_AA}) are in general given by different ${\bm R}-{\bm R'}$, e.g. ${\bm R}-{\bm R'}\pm \bm{a}_1$, ${\bm R}-{\bm R'}\pm \bm{a}_2$, etc. Here $\bm{a}_1$ and $\bm{a}_2$ are the primitive vectors of the Bravais lattice . However in the long distance limit, all the decay  factors can be replaced by $1/|{\bm R}-{\bm R'}| ^3$. 

Treating the cosine is a bit more delicate as $\cos\bigl(2 {\mathbf k}_D\cdot ( {\bm R}-{\bm R'} )\bigr)$ is not a smooth function. It is straightforward to show that $\cos\left(2 {\mathbf k}_D\cdot {\bm R} \right)$ is either $-1/2$ or $1$. If we decompose $\bm{R}$ into the primitive vectors 
\begin{align}
\bm{a}_1 &=\left(3/2,\sqrt{3}/2 \right), \\
\bm{a}_2 &=\left(0,-\sqrt{3}\right),\\
\bm{R} &= m \bm{a}_1 + n \bm{a}_2,
\end{align}
together with
\begin{equation}
\bm{k}_D =\left(0,\frac{4\pi}{3\sqrt{3}}\right),
\end{equation}
one finds
\begin{equation}
\cos\left(2 {\mathbf k}_D\cdot {\bm R} \right) = \cos\left(4\pi(m+n)/3\right).
\end{equation}
 However as we explain this complication does not come into play and the RKKY exchange between the plaquette impurities we have considered is always antiferromagnetic. 

\begin{figure}
\includegraphics{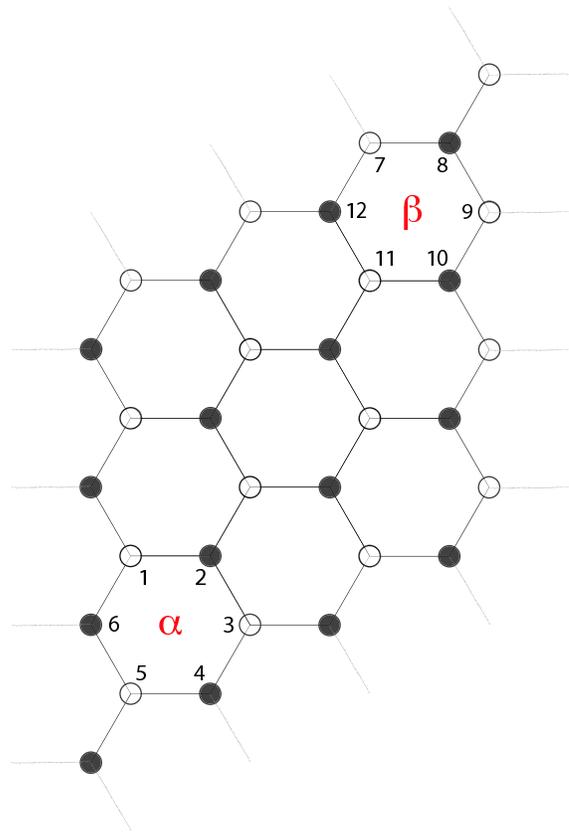}
\caption{\label{cos} (Color online) The honeycomb lattice. ``A" sublattice is denoted by odd (empty) sites. Plaquettes $\alpha$, $\beta$, and the sites surrounding them are labeled. Fixing i on any site and running $\rm j\in A$ (around a plaquette) results in $\{0,1,2\}$ patterns for $O_{\rm ij} = m_{\rm i}+n_{\rm i}-m_{\rm j}-n_{\rm j}  \ (\rm mod~3) $. The same set is obtained by scanning $\rm j\in B$. Thus this procedure results in the set $\{-1/2,-1/2,1\}$  for $\cos(2\bm{k}_D\cdot \bm{R}_{\rm ij})$ and $\cos(\bm{k}_D\cdot \bm{R}_{\rm ij})$. }
\end{figure}

To see this one has to label the sites around the two hexagons in terms of $(m,n)$ pairs. It is straightforward to see that the set of 
\begin{equation} O = m+n-m'-n' \ (\rm mod~3)\end{equation}
for AA and BB sublattices is the same as the set for opposite AB sublattices, thus antiferromagnetism prevails. It can is also found (see Fig.~\ref{cos}) that these sets will be grouped to $\{0,1,2\}$. Therefore the cosine contributions of Eq.~(\ref{EQ:J_AB}) and Eq.~(\ref{EQ:J_AA}), when summed over the sites around two plaquettes,  vanish: to the leading order $J^{\alpha \beta} (\bm{R}-\bm{R}')$ is given by

\begin{equation} \label{EQ:J_alphabeta}
J^{\alpha \beta} (\bm{R}-\bm{R}') \approx \frac{9\mathfrak{j}^2}{16\pi }\frac{1}{ |{\bm R}-{\bm R'}| ^3}.
\end{equation}

To summarize, the RKKY exchange for plaquette impurities (in the long distance limit) is always antiferromagnetic. Since the center of the hexagons form a triangular lattice, the  RKKY exchange for the widely separated plaquette impurities is frustrated.

We also mention the results for the RKKY interaction between a plaquette impurity and a site impurity. Again the $\cos\left(2 {\mathbf k}_D\cdot ( {\bm R}-{\bm R'} )\right)$, present in the RKKY for site impurities, when summed over the plaquette, vanishes. The final result is given by
\begin{equation} \label{EQ:J_alphaA}
J^{\alpha}_{A} (\bm{R}-\bm{R}') \approx \frac{3\mathfrak{j}^2}{32\pi }\frac{1}{ |{\bm R}-{\bm R'}| ^3}.
\end{equation}
The same result is obtained when the site impurity is on a B sublattice.

We finally extend  the incoherent Kondo coupling of plaquette impurities with the conduction electron given by Eq.~(\ref{EQ:general_pert}) to the coherent one. The coherent Kondo perturbation is more physical. It can be justified by going back to the origin of the Kondo model, i.e. the Anderson model. In the coherent Kondo coupling the $S^{\alpha}$ has a Kondo interaction with a coherent sum of conduction electron spin. The Kondo perturbation is then given by
\begin{equation} \label{EQ:coherent}
\delta \hat{H} = \mathfrak{j}~ \bm{S}^{\alpha}\cdot \bm{s}_{\alpha } \ \ \ (|\mathfrak{j}| \ll t),
\end{equation}
where $\bm{s}_{\alpha }$ is a coherent sum of the conduction electron spins around the plaquette $\alpha$, eg. $s_{\alpha }^+$ is given by
\<
s_{\alpha }^+ = \frac{1}{6} \sum_{\rm i, j \in \mathcal{P}^\alpha} \hat{c}_{\rm i \uparrow}^\dagger \hat{c}_{\rm j \downarrow},
\>
and $\mathcal{P}^\alpha$ is the set of sites surrounding the plaquette $\alpha$. The susceptibility needed in these calculations carries four sublattice indices:
\begin{equation} \label{EQ:chi}
\chi^{\rm{DD'}}_{\rm ikjl}(q)=\int\frac{d^3 k}{(2\pi)^3}\ \ 
{\rm ^i _j}\ \parbox{11mm}{
\begin{fmfgraph*}(10,6)
\fmfleft{i}
\fmfright{o}
\fmf{dbl_plain_arrow,label=$k+q; \rm{D}$,right}{o,i}
\fmf{dbl_plain_arrow,label=$k; \rm{D'} $,right}{i,o}
\fmfv{
decor.shape=cross,decor.filled=full,decor.size=2thick}{i}
\fmfv{
decor.shape=cross,decor.filled=full,decor.size=2thick}{o}
\end{fmfgraph*}
}{\rm ^k _l}\ ,
\end{equation}
since in calculating  $\left< s_{\alpha }^+ s_{\beta }^-\right>$, 4 sites $\{\rm i,j,k,l\}$ are involved. Here $\rm i,j \in \mathcal{P}^\alpha$ and $\rm k,l \in \mathcal{P}^\beta$. We avoided using a separate notation for sublattice indices in $\chi^{\rm{DD'}}_{\rm ikjl}(q)$, e.g. ``i" in $\chi^{\rm{DD'}}_{\rm ikjl}(q)$ should be replaced with A if $\rm i \in A$, etc. $J^{\alpha \beta}_*$ is then given by:
\begin{widetext}
\begin{equation}\label{EQ:J*}
J^{\alpha \beta}_* \approx -\mathfrak{j}^2 \sum_{\rm i, j \in \mathcal{P}^\alpha} \sum_{\rm k , \rm l \in \mathcal{P}^\beta} \sum_{\rm{D,D'}} \int \frac{d^2{\bm q}}{(2\pi)^2}e^{i(\bm{q}+D \bm{k}_D) \cdot({\bm R}_{\rm i}-{\bm R}_{\rm k})}e^{iD' \bm{k}_D \cdot({\bm R}_{\rm l}-{\bm R}_{\rm j})}   \chi^{\rm{DD'}}_{\rm ikjl}(q_0=0,{\bm q}) \mathcal{C}_\Lambda(|\bm{q}|).
\end{equation}
\end{widetext}
``*" is used to make the distinction between the coherent RKKY and the incoherent RKKY of Eq.~(\ref{EQ:J_alphabeta}).

It is straightforward  to find that:
\begin{equation}\label{*=0}
\sum_{\rm i, j \in \mathcal{P}^\alpha}^{\ \ \ \ \prime} \sum_{\rm k , \rm l \in \mathcal{P}^\beta}^{\ \ \ \ \prime}  \sum_{\rm{D,D'}} e^{iD \bm{k}_D \cdot({\bm R}_{\rm i}-{\bm R}_{\rm k})}e^{iD' \bm{k}_D \cdot({\bm R}_{\rm l}-{\bm R}_{\rm j})} =0,
\end{equation}
where $\sum^{\prime}$ is to denote that in summing over indices, each site index is restricted to be in a \textit{single} sublattice. This is just an immediate result of the observation we made in the previous discussions and is explained briefly in the caption of Fig.~\ref{cos}. Since 
\begin{equation}
\int \frac{d^2{\bm q}}{(2\pi)^2}e^{i\bm{q} \cdot({\bm R}_{\rm i}-{\bm R}_{\rm k})} \chi^{\rm{DD'}}_{\rm ikjl}(q_0=0,{\bm q}) \mathcal{C}_\Lambda(|\bm{q}|),
\end{equation}
is independent  of $D$ and $D'$, Eq.~(\ref{*=0}) implies that the $1/|\bm{R}-\bm{R}'|^3$ dependance of $J^{\alpha \beta}_* $ vanishes:
\begin{equation} \label{Jcohere}
J^{\alpha \beta}_* (\bm{R}-\bm{R}') = 0 + \mathcal{O}(1/|\bm{R}-\bm{R}'|^4).
\end{equation}

Note that the \textit{incoherent} Kondo perturbation corresponds to $\rm i=j$ and $\rm k=l$. The sum in Eq.~(\ref{*=0}), with this further constraint, does not factor out and it does not vanish [See Eq.~(\ref{EQ:J_alphabeta})].



\section{conclusions}

We first proved that the particle-hole symmetry for bipartite lattices determines the sign of RKKY interaction between site impurities on all length scales. Secondly, the nature of the singularity of the spin susceptibility in graphene invalidates the use of sharp cutoff. In the sharp cutoff scheme, the main contribution for the RKKY interaction comes from the large momenta, thus invalidating the low energy theory. 

We then studied the RKKY between plaquette impurities and also between a plaquette impurity and a site impurity. We first considered the simpler case of having an incoherent Kondo perturbation. For these incoherent perturbations, the RKKY involving   plaquette impurities (in the long distance limit) always ends up to be \textit{antiferromagnetic}.  This is coming from the fact that, in mediating RKKY the contributions of electrons on opposite sublattices dominate  the contributions from the same sublattices [Eq.~(\ref{EQ:J_AB}) vs. Eq.~(\ref{EQ:J_AA})]. If they had equal strength, the $1/|{\bm R}-{\bm R'}| ^3$ dependance of the RKKY involving plaquette impurities would have vanished. 

We then focused on plaquette impurities in the \textit{coherent}  Kondo interaction regime. We found that the $1/R^3$ algebraic tail of the RKKY, present in all other situations we considered, vanishes. More work needs to be done for finding the leading contribution for the RKKY in this case. 

In terms of the \textit{magnitude} of the RKKY in the long distance limit, our results can be summarized symbolically as 
\begin{equation}
|J^{\alpha\beta}|>|J^{\alpha}_{\rm A}|>|J_{\rm AB}|>|J_{\rm AA}| \gg |J_*^{\alpha \beta}|.
\end{equation}

There had been other attempts in calculating RKKY in graphene.\cite{VLSG05, DLB06} They also observed a $1/R^3$ dependance for the RKKY.  However it was claimed that the RKKY is \textit{always} ferromagnetic due to graphene's ``semimetalic properties". Based on our result, this \textit{may} only happen for the plaquette impurities with coherent Kondo interactions [See Eq.~(\ref{Jcohere})], and in that case the RKKY decay is faster than  $1/R^3$.

Finally we mention the implications  of this paper for the Kondo lattice model on the honeycomb lattice. We started the study of the Kondo-Heisenberg model on the honeycomb lattice\cite{saeed07}  with the hope of finding the first examples of algebraic spin liquid (ASL) phase in the presence of a semimetal; or  a Kondo insulator--N\'{e}el deconfined quantum critical point.\cite{Senthiletal}
Based on the result of Sec.~\ref{SEC:lattice}, $J_H>0$ {\it will be generated} in the small $J_K$ limit of the Kondo lattice model
\begin{equation} \label{EQ:kl}
\hat{H} = -t \sum_{\left<\rm ij\right>,a} \left( \hat{c}_{\rm i}^{a\dagger} \hat{c}_{\rm j}^{a} + {\rm H.c.} \right) + J_K \sum_{\rm i} \bm{s}_{\rm i}\cdot \bm{S}_{\rm i},
\end{equation}
and in this regard the Kondo lattice model transforms to the model we studied in Ref.~\onlinecite{saeed07}. 


\section{Acknowledgments}
I thank Patrick Lee for very useful discussions I have had with him about this and related issues and for encouraging me to write this paper. I am grateful for the conversations I had with Xiao-Gang Wen, Michael Hermele, Ying Ran, and Cody Nave.  I am also grateful for the email exchanges I had with F. Guinea after this paper appeared on arXiv. The consideration of coherent Kondo couplings for plaquette impurities was especially motivated by his comment.

\end{fmffile}

\appendix
\section{Inetgral Table \label{APP:integrals}}
\begin{align}
&\int \frac{ d^3k}{(2\pi)^3} \frac{k_{\mu}(k+q)_{\nu}}{k^2(k+q)^2}=\frac{-1}{64|q|}\left(q_\mu q_\nu + q^2 \delta_{\mu\nu}\right)+\cdots,\\
&\int_0^{2\pi} d\theta~e^{ix\cos\theta}e^{in\theta} = 2\pi i^{n}J_n(x),\\
\lim_{\alpha\rightarrow\infty} &\int_0^{\infty} dx~ x^2e^{-x/\alpha}J_0(x)= -1,\\
\lim_{\alpha\rightarrow\infty} &\int_0^{\infty} dx~ x^2e^{-x/\alpha}J_2(x)= +3,\\
\lim_{\alpha\rightarrow\infty} &\int_0^{\infty} dx~ x^2e^{-x^2/\alpha^2}J_0(x)= -1,\\
\lim_{\alpha\rightarrow\infty} &\int_0^{\infty} dx~ x^2e^{-x^2/\alpha^2}J_2(x)= +3.
\end{align}
The ellipsis in the first equation is to denote the non-universal pieces of the integral. They cause either exponential decay, or faster algebraic decays, in the continuum limit.

\end{document}